\documentclass[10pt]{iopart}
\usepackage{a4,epsfig}
\usepackage{url}
\usepackage[dvipsnames]{xcolor}
\usepackage{lineno}
\usepackage[normalem]{ulem}
\usepackage{fancyhdr}
\usepackage{graphicx}
\usepackage{float}
\usepackage{multirow}
\expandafter\let\csname equation*\endcsname\relax
\expandafter\let\csname endequation*\endcsname\relax
\usepackage{amsfonts}
\usepackage{amssymb}
\usepackage{amsmath}
\usepackage{hyperref}
\usepackage{subfigure}
\usepackage{lineno}
\usepackage{upgreek}
\usepackage{comment}
\usepackage{caption}

\begin{document}
\title{Newtonian calibrator tests during the Virgo O3 data taking}

\author{D. Estevez, B. Mours, T. Pradier}

\address{Institut Pluridisciplinaire Hubert CURIEN, 23 rue du loess - BP28 67037 Strasbourg cedex 2, France}

\ead{dimitri.estevez@iphc.cnrs.fr}

\begin{abstract}
The gravitational-wave detectors outputs from the LIGO and Virgo collaborations have been a source of scientific results of prime importance in various domains such as astrophysics, cosmology or fundamental physics. With the upgrades of the detectors and their improved sensitivities, new challenges are set for these instruments calibration. A calibration method based on the local variations of the Newtonian gravitational field could be the next absolute reference of calibration for the interferometers network.
We report new tests of Newtonian calibrators (NCal) on the Advanced Virgo detector performed during the LIGO-Virgo observing run O3. The NCal-induced strain on a mirror of the interferometer has been computed both using analytical calculations and numerical simulations with results in very good agreement. The strains given by the numerical model have been used to analyse the data of the NCals and have been compared to the reference method of calibration using photon radiation pressure. New methods to measure the NCal to mirror distance and the NCal offset with respect to the plane of the interferometer using two NCals are also presented. They are used to correct the NCal data which improves the agreement with the current Advanced Virgo reference of calibration.    
\end{abstract}
\maketitle

\section{Introduction} \label{introduction}
Accurate calibration of  gravitational-wave detectors is critical for proper sky localization, 
testing waveforms and General Relativity, or extracting measurements like the Hubble constant \cite{GWTC} \cite{testingGR} \cite{Hubble}. 
The requirements will become even more severe with the planned sensitivity improvements \cite{ObsSce}. 
Calibration using a locally induced variation of the gravity field, using rotating objects is an old idea \cite{Sinsky} \cite{Hirakawa80} \cite{Kuroda85} \cite{Mio87} \cite{Explorer91} \cite{Explorer98} \cite{Hild07} \cite{Matone07}.
This kind of system, named Newtonian Calibrator (NCal) or sometimes Gravity field Calibrator (GCal),  
has been first tested on a large interferometer (Virgo) during the so called O2 data taking \cite{NCal_virgo}.
Since then, new prototypes have been built and tested at LIGO \cite{NCal_ligo}, KAGRA \cite{NCal_KAGRA} and Virgo \cite{NCal_virgo2}.
   
The calibration of a gravitational-wave detector requires the injection of a well known signal \cite{virgocal} \cite{ligocal}.
This means that we need to build a device which will induce a known displacement of a test mass of the interferometer.
We also need to accurately predict the injected signal. The typical accuracy of the Virgo calibration reference, called the Photon Calibrators (PCal), is $1.4\%$ for the third observing run O3 (April 2019, to March 2020) \cite{AdVPCal}.
This value sets the current level of accuracy to reach with the NCal if one wants to use it as a calibration tool on Virgo. 
Following the first tests made on Virgo during O2, 
new improved NCals have been built for the O3 data taking.
The modeling of the expected signal has also been developed, 
both on the analytical side, with a more complete formalism, 
and on the numerical side, with a new detailed finite element analysis program named FROMAGE.

This paper reports these new developments and the achieved results.
It starts by the presentation of the Virgo NCal layout used during O3. 
Next, the analytical and numerical modeling are detailed.
The description of the collected data and first checks follow. 
They are used to derive the NCals positions relative to the mirror location. 
Using this corrected positions, a check of the $h(t)$ reconstruction is performed 
and compared to the check made with the reference calibration technique for Virgo, the so called photon calibrator.
Finally, the future evolution of the Virgo NCal system is briefly discussed.

\section{Virgo NCal setup for O3} \label{setup}
Following the NCal tests made during the Virgo Observing run O2, a new NCal system was built for O3. The goal was to achieve a higher frequency signal, with a quieter system and reduced systematic uncertainties. 

\begin{figure} [h]
\centerline{\includegraphics[width=1\textwidth]{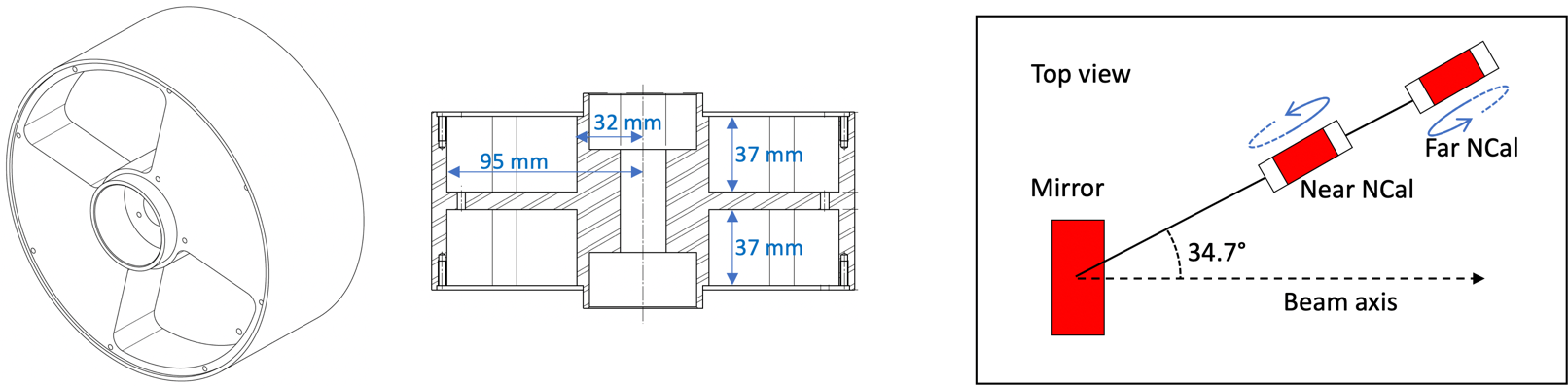}}
\caption{Left: 3d view of the O3-NCal rotor, center: cross section of the rotor; right: sketch of the NCals locations relative to the mirror}
\label{fig:NCal-sketch}
\end{figure}

A more compact rotor was designed, as described on figure \ref{fig:NCal-sketch}. 
It is still made of aluminum for easiness of machining, but with two 90 degrees sectors for optimal strength. 
The external rotor diameter is 205~mm. 
Two cavities are made on each side of the rotor, with a nominal inner radius of 32~mm, outer radius of 95 mm and 37 mm depth. 

One of the rotor was measured with a 3d system,
and a systematic deviation of the cavities depth of about 0.1 mm was observed on both sides of the rotor. 
Therefore, when computing the expected displacement, the thickness of the "active" part, i.e. the $90^{\circ}$ sectors made of aluminum, of the rotor is taken as $b = 73.8$~mm.
The uncertainty on the rotor thickness is set to this correction as a conservative approach.
The outer radius of the cavities was found to be systematically larger by 0.045 mm. 
Again, when computing the radius of the active part of the rotor we used this measure value of $r_{\textrm{max}} = 95.045$~mm. 
The uncertainty on $r_{\textrm{max}}$ is this correction. 
The rotors are made of aluminum 7075, whose expected density is $\rho$ = 2805~kg/m$^3$ with a typical uncertainty of 5~kg/m$^3$, larger than the air density that is neglected.
A global test was made by measuring the weight of the rotor: 4.837~kg, which compares very well with the expected CAD value of 4.838~kg, when using the measured rotor parameters.
This agreement is an indication that the uncertainties on the rotor parameters are not underestimated.

In first approximation, the amplitude of the calibration signal injected by the NCal is proportional to the rotor thickness, density and to the fourth power of the rotor external radius (see equation \ref{eq:8}). 
Therefore, we can derive the uncertainty on this amplitude from the uncertainties on the main rotor parameters (see table \ref{table:rotor-uncertainties}). 

\begin{table}[h!]
\begin{center}
\begin{tabular}{|c|c|c|c|c|}
\hline
\multicolumn{3}{|c|}{Rotor parameter} & \multicolumn{2}{c|}{NCal signal uncertainty }\\
\hline
name & value & uncertainty     & formula & value (\%) \\
\hline
Density $\rho $ (kg/m$^3 $) & 2805 & 5 & $\delta \rho/\rho$ & 0.18 \\
Thickness $b$ (mm) & 74 & 0.2 & $\delta b/b$ & 0.27 \\
$r_{\textrm{max}}$ (mm) & 95 & 0.1 & $4 \delta r_{\textrm{max}}/r_{\textrm{max}}$ & 0.42 \\
\hline
\multicolumn{4}{|c|}{Total uncertainty from the rotor (quadratic sum )} & 0.53\\
\hline
\end{tabular}
\caption{Uncertainties on the amplitude of the calibration signal from the rotor geometry}
\label{table:rotor-uncertainties}
\end{center}
\end{table}

Both sides of the rotor are covered by plastic covers to avoid air motion.
The rotor is enclosed in an aluminum housing for safety reason. 
Instead of driving the rotor with pulleys and belt like for the first prototype, 
a frameless motor is directly installed on the rotor axis, 
reducing significantly the vibrations generated by the rotor rotation. 
The size of the bearing used was reduced to minimize the losses, but their choice and mounting turned out to be difficult, with sometimes not fully reproducible results for the maximum rotation speed.
The frequency of the rotor as well as the phase information is retrieved using a photodiode sensing and a LED located on the other side of the rotor through a hole made in the rotor.

Instead of having the NCal just sitting on the base of the vacuum chamber like during the O2 test, 
we hung the NCal from the structure surrounding the vacuum chamber with a soft suspension.
The knowledge of the mirror to NCal distance is the most difficult part. 
A direct measurement is impossible, since the mirror is inside the vacuum chamber, without direct view from the NCal position. 
Furthermore, the suspended solution used to filter the NCal residual vibrations makes this distance not well known. 
Therefore, we built two NCal devices that, after some tests, were installed on the same side of the vacuum chamber. 
They are mounted on a rigid rod, with a relative distance of 680~mm. 
The expected mirror to NCal distance was 1.267~m for the NCal closer to the mirror, or near NCal, and 1.947~m for the far NCal.  
The angle between the beam axis and the NCals to mirror axis was the same for both NCal: 34.71~degrees.

The NCals are located on the plane of the interferometer, which 
significantly reduces the effect of the vertical positioning uncertainty. 
The vertical positions of the two NCals have been measured to be the same within 0.5 mm. 
The uncertainty on the difference of the phase origins of the two NCals is less than 0.5 mrad, 
and 1 mrad for each of the individual absolute phase origin.
The far NCal was flipped compared to the near NCal, meaning that they were rotating in opposite directions, a feature that allow us to extract the vertical position as described in a following section.

\section{Improved analytical model }
\begin{figure}[h!] 
 \begin{center}
 	\captionsetup{justification=justified}
    \subfigure[Top view]{
	   \includegraphics[trim={1.2cm 0cm 1.2cm 1cm},clip,scale=0.4]{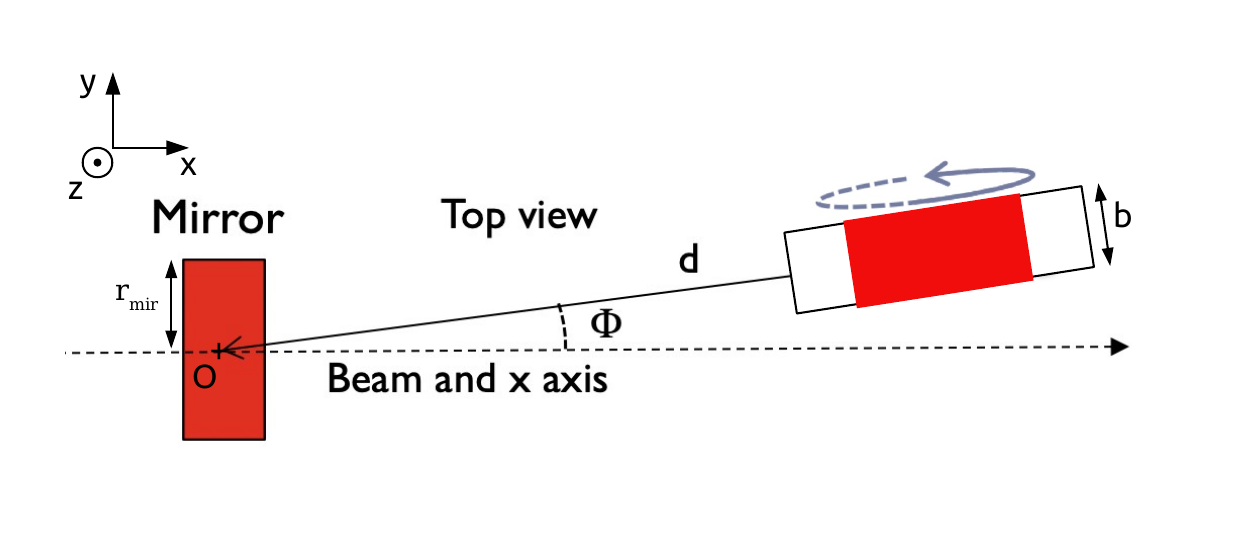} 
	   \label{fig:sketchPhi} }
    \subfigure[Side view]{
       \includegraphics[trim={1.2cm 1cm 1.2cm 1cm},clip,scale = 0.4]{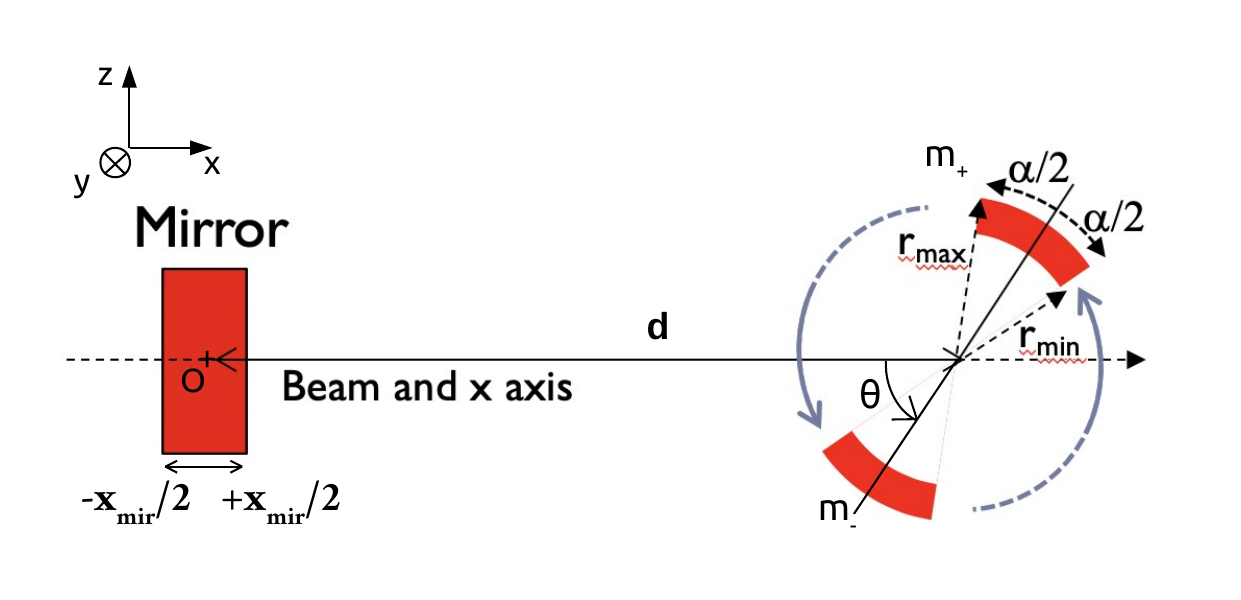}
       \label{fig:sketchMirRot} }
	\caption{Schematic of the NCal rotor position with respect to an end mirror of Advanced Virgo. A top view is shown in (a) and a side view in (b).}
	\label{fig:sketchNCal}
	\end{center}
\end{figure}

The point-mass analytical calculation of the quadrupole force of the Virgo NCal has been refined by including an extended rotor and an extended mirror. Figure \ref{fig:sketchPhi} and \ref{fig:sketchMirRot} show respectively a top view and a side view of the considered NCal configuration.
\\The reference frame is $(O,x,y,z)$ with the center of the mirror in $O$. The mirror is assumed to be a full cylinder of radius $r_{mir}$ and thickness $x_{mir}$. The cylindrical coordinates of a point of the mirror in the reference frame are: $$(x';r'\cos(\beta);r'\sin(\beta))$$
where $x'\in [-x_{mir}/2,x_{mir}/2]$, $r'\in [0,r_{mir}]$ and $\beta \in [0,2\pi]$.\\
Here we assume for simplicity that the rotor axis is in the plane of the interferometer ($z=0$) but a detailed calculation when $z \neq 0 $ is given in appendix \ref{app:analytical}. The coordinates of a point of the rotor sector masses $(+,-)$ in the reference frame are:
$$\begin{pmatrix}d\cos(\phi)\pm r\cos(\theta+\psi)\cos(\phi) - b'\sin(\phi) \\ d\sin(\phi)\pm r\cos(\theta+\psi)\sin(\phi) + b'\cos(\phi) \\ \pm r\sin(\theta+\psi) \end{pmatrix}$$
where $d$ is the distance from the center of the rotor to the center of the mirror, $\phi$ is the angle between the beam axis of the interferometer and the rotor-to-mirror axis, $\theta$ is the rotation angle of the masses around the rotor axis, $r \in [r_{min},r_{max}]$ with $r_{min}$ and $r_{max}$ the minimum and maximum values of the radial dimension of a rotor sector, $b' \in [-b/2,b/2]$ with $b$ the thickness of the rotor, and $\psi \in [-\alpha/2,\alpha/2]$ with $\alpha$ the opening angle of a rotor sector.
Defining the small quantities $\epsilon = \frac{r}{d}$, $\epsilon' = \frac{r'}{d}$, $\epsilon'' = \frac{x'}{d}$ and $\epsilon''' = \frac{b'}{d}$, the longitudinal force induced by the rotor masses $(+,-)$ is:
\begin{align}
F_{\pm,x} & = \frac{GmM}{d^2}(\cos(\phi)\pm\epsilon \cos(\theta+\psi)\cos(\phi)-\epsilon'''\sin(\phi)-\epsilon'')\Big[1+X_{\pm}\Big]^{-3/2}
\label{eq:1}
\end{align}
where $G$ is the gravitational constant, $m$ is the mass of the masses $(+,-)$, $M$ is the mass of the mirror and:
\begin{align}
X_{\pm} & = \epsilon^2+\epsilon'^2+\epsilon''^2+\epsilon'''^2\pm2\epsilon\cos(\theta+\psi)-2\epsilon''\cos(\phi)(1\pm\epsilon\cos(\theta+\psi))\nonumber \\
&~~-2\epsilon'\sin(\phi)\cos(\beta)(1\pm\epsilon\cos(\theta+\psi))\mp 2\epsilon\epsilon'\sin(\beta)\sin(\theta+\psi)) \nonumber \\
&~~+2\epsilon'''(\epsilon''\sin(\phi)-\epsilon'\cos(\phi)\cos(\beta))
\label{eq:2}
\end{align}
The expression from equation \ref{eq:2} can be expanded at the fourth order in $X_{\pm}$ using:
\begin{equation}
 (1+X_{\pm})^{-3/2} \approx\Big(1-\frac{3}{2}X_{\pm}+\frac{15}{8}X_{\pm}^2-\frac{35}{16}X_{\pm}^3+\frac{315}{128}X_{\pm}^4\Big)
\label{eq:3}
\end{equation}
Keeping only the time dependent terms at twice the rotor frequency up to the fourth order and summing the longitudinal force of both rotor sectors, the total longitudinal force $F_x$ is:
\begin{align}
F_{x} & \approx \frac{9GmMr^2}{2d^4}\cos(\phi)\Big[1+\frac{25}{36}\epsilon^2+\Big(\frac{45}{2}\sin^2(\phi)\cos^2(\beta)-\frac{15}{9}\sin^2(\beta)-\frac{25}{6}\Big)\epsilon'^2 \nonumber \\
&~~~~~~~~~~~~~~~~~~~~~~~~~~~+\Big(\frac{45}{2}\cos^2(\phi)-\frac{25}{9}\Big)\epsilon''^2-\frac{25}{6}\epsilon'''^2\Big]\cos(2(\theta+\psi))
\label{eq:4}
\end{align}
where the odd terms in $\cos(\beta)$, $\sin(\beta)$, $\epsilon''$ and $\epsilon'''$ have been omitted for clarity reasons since they cancel when integrating over a small volume element of the mirror in next equations.\\ 
Considering a small element of the rotor and of the mirror respectively with a mass $\text{d}m_{rot}~ =~ \rho_{rot} ~ r \text{d}r ~\text{d}\psi ~\text{d}b' $ and $\text{d}m_{mir} = \rho_{mir} ~ r' \text{d}r' ~\text{d}\beta ~\text{d}x' $, the mass of a rotor sector and of the mirror are:
\begin{equation}
m=\rho_{rot}\int_{r_{min}}^{r_{max}} \int_{-\frac{\alpha}{2}}^{\frac{\alpha}{2}} \int_{-\frac{b}{2}}^{\frac{b}{2}} r~dr~d\psi~db'
\label{eq:5}
\end{equation}
\begin{equation}
M=\rho_{mir}\int_{0}^{r_{mir}} \int_{0}^{2\pi} \int_{-\frac{x_{mir}}{2}}^{\frac{x_{mir}}{2}} r'~dr'~d\beta~dx'
\label{eq:6}
\end{equation}
It is then possible to obtain the total longitudinal force for an extended rotor and an extended mirror by integrating equation \ref{eq:4} using equations \ref{eq:5} and \ref{eq:6}. Since the amplitude of an Advanced Virgo mirror motion follows a free test mass response above the resonance frequency of the suspension ($f\gg 0.6~$Hz), it is expressed as:
\begin{equation}
a(f_{2rot}) = \frac{|F_x|}{M(2\pi f_{2rot})^2}
\label{eq:7}
\end{equation}
where $f_{2rot} = 2 f_{rot}$ with $f_{rot}$ the rotor frequency.\\
The resulting amplitude is thus:
\begin{align}
a(f_{2rot}) & = \frac{9 G \rho_{rot}~ b~\sin(\alpha)(r_{max}^{4}-r_{min}^{4})}{32\pi^2 f_{2rot}^{2}d^4}\cos(\phi)\Big[1+\frac{25}{54d^2}\frac{(r_{max}^{6}-r_{min}^{6})}{(r_{max}^{4}-r_{min}^{4})}+\Big(\frac{45}{8}\sin^2(\phi)-\frac{5}{2}\Big)\Big(\frac{r_{mir}}{d}\Big)^2 \nonumber \\
&~~~~~~~~~~~~~~~~~~~~~~~~~~~~~~~~~~~~~~~~~~~~~~~~~~~~~+\Big(\frac{15}{8}\cos^2(\phi)-\frac{25}{24}\Big)\Big(\frac{x_{mir}}{d}\Big)^2-\frac{25}{72}\Big(\frac{b}{d}\Big)^2\Big]
\label{eq:8}
\end{align}
The opening angle $\alpha$ has been set to $90^{\circ}$ for the Virgo NCal since it maximizes the induced signal. We can also define a strain which is the amplitude normalized by the interferometer arm length $L=3~$km for Virgo:
\begin{equation}
h(f_{2rot}) = \frac{a(f_{2rot})}{L}
\label{eq:9}
\end{equation}
For instance, computing the amplitude of the mirror motion using the Virgo O3 NCal near parameters one can find $1.007136$e-$14~f_{2rot}^{-2}~$m.Hz$^2$ or in strain unit $3.35712$e-$18~f_{2rot}^{-2}~$Hz$^2$.

\section{Finite element modeling with FROMAGE} \label{fromage}
The analytical formula for the NCal works well when the rotor and mirror geometry can be described in simple volume elements such as cylinders, sectors, cuboids etc... However, in reality the objects we consider have some imperfections that cannot be modeled in a simple analytical manner.

\begin{figure}[!h]
	\center
	\includegraphics[trim={0cm 1cm 0cm 1cm},clip,scale=0.5]{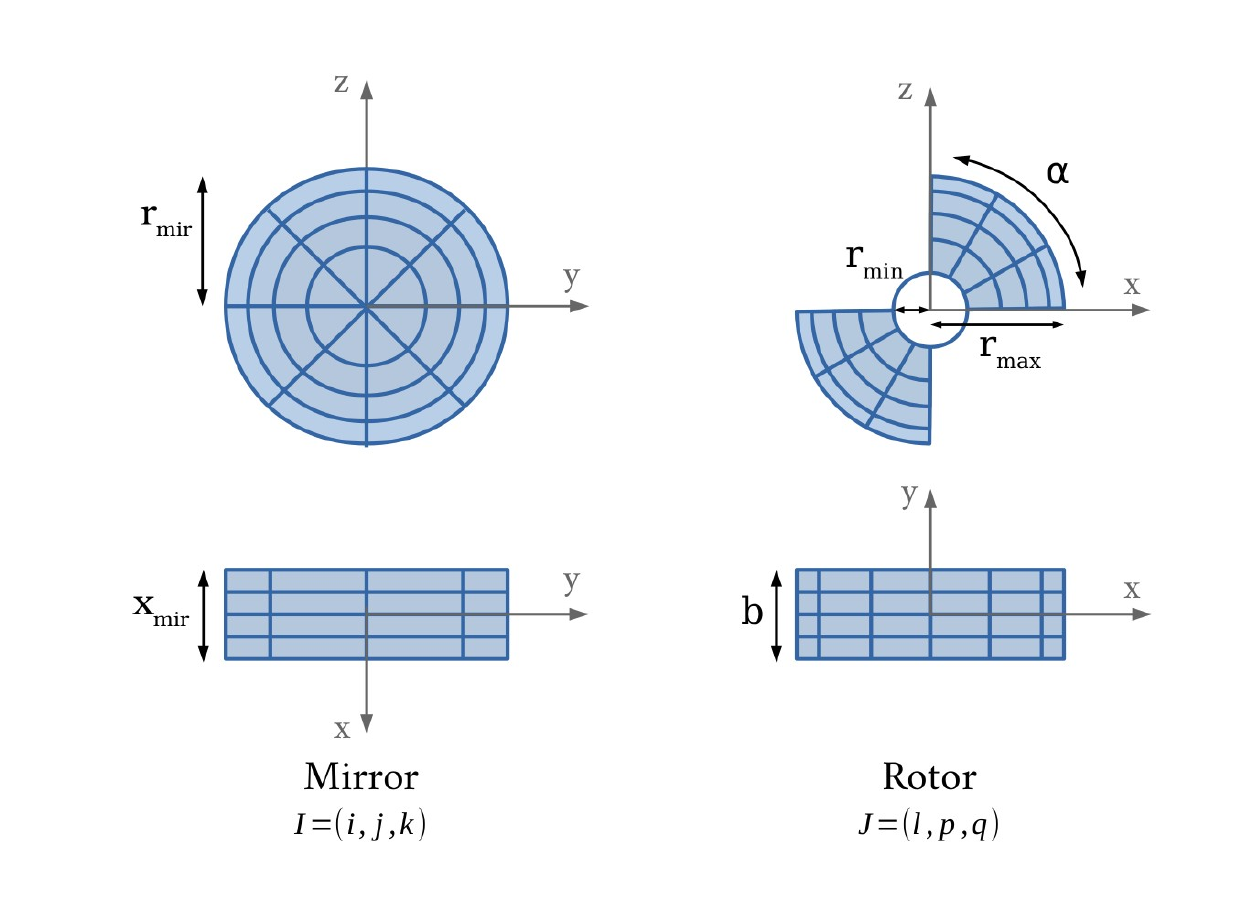} 
	\caption{Discretization of the mirror and the rotor respectively in $i\times j\times k$ and $l\times p\times q$ elements.}
	\label{fig:FEM_mirrorandrotor}
\end{figure}

A straightforward way to compute the Newtonian gravitational force for the NCal with an arbitrary geometry is to use a finite element analysis by discretizing the rotor and the mirror into small 3D elements with simple geometrical forms as shown in figure \ref{fig:FEM_mirrorandrotor}. This is the aim of the simulation tool, called FROMAGE\footnote{Finite element analysis of ROtating MAsses for Gravitational Effects} \cite{froma}, we developed.

Computing and adding the gravitational force of each rotor element applied to each mirror element, the total force on the mirror can be derived:
\begin{equation}
\vec{F} = G \sum_{I}\sum_{J}\frac{m_{mir,I}~m_{rot,J}}{d_{I,J}^2}\vec{u}_{I,J}
\label{eq:10}
\end{equation}
with $I=(i,j,k)$ and $J=(l,p,q)$ the indexes of the mirror and the rotor elements respectively, $d_{I,J}$ the distance between the mirror element $I$ and the rotor element $J$ and $\vec{u}_{I,J}$ a unit vector along $d_{I,J}$.\\
Equation \ref{eq:10} can then be projected onto the $x$ axis to obtain the NCal-induced longitudinal force on the mirror:
\begin{equation}
F_{x} =   G \sum_{I}\sum_{J} \frac{m_{mir,I}~m_{rot,J}(x_{rot,J}-x_{mir,I})}{d_{I,J}^3}
\label{eq:11}
\end{equation}
In the case of the Virgo NCal this force is expected to be $\pi-$periodic since the two sectors should only induce a quadrupole component (i.e at $2f_{rot}$). However, in order to properly extract any contribution of harmonics in the force, we expand the numerical force in Fourier series such that:
\begin{equation}
F_{x}(\theta) = \frac{1}{N}\Big(C_0 + 2\sum_{k=1}^{N-1} |C_k| \cos(k\theta + \Phi_k)\Big)
\label{eq:12}
\end{equation} 
with $N$ the maximum order to which the signal is expanded, $C_{k}$ the Fourier coefficients of the force and $\Phi_{k}$ the phase of the Fourier coefficients. The results reported further in this paper for the Virgo NCal are indeed the $|C_{2}|$ values extracted from equation \ref{eq:12}.

\begin{figure}[!h]
	\center
	\captionsetup{justification=justified}
	\includegraphics[trim={0cm 1cm 0cm 1cm},clip,scale=0.4]{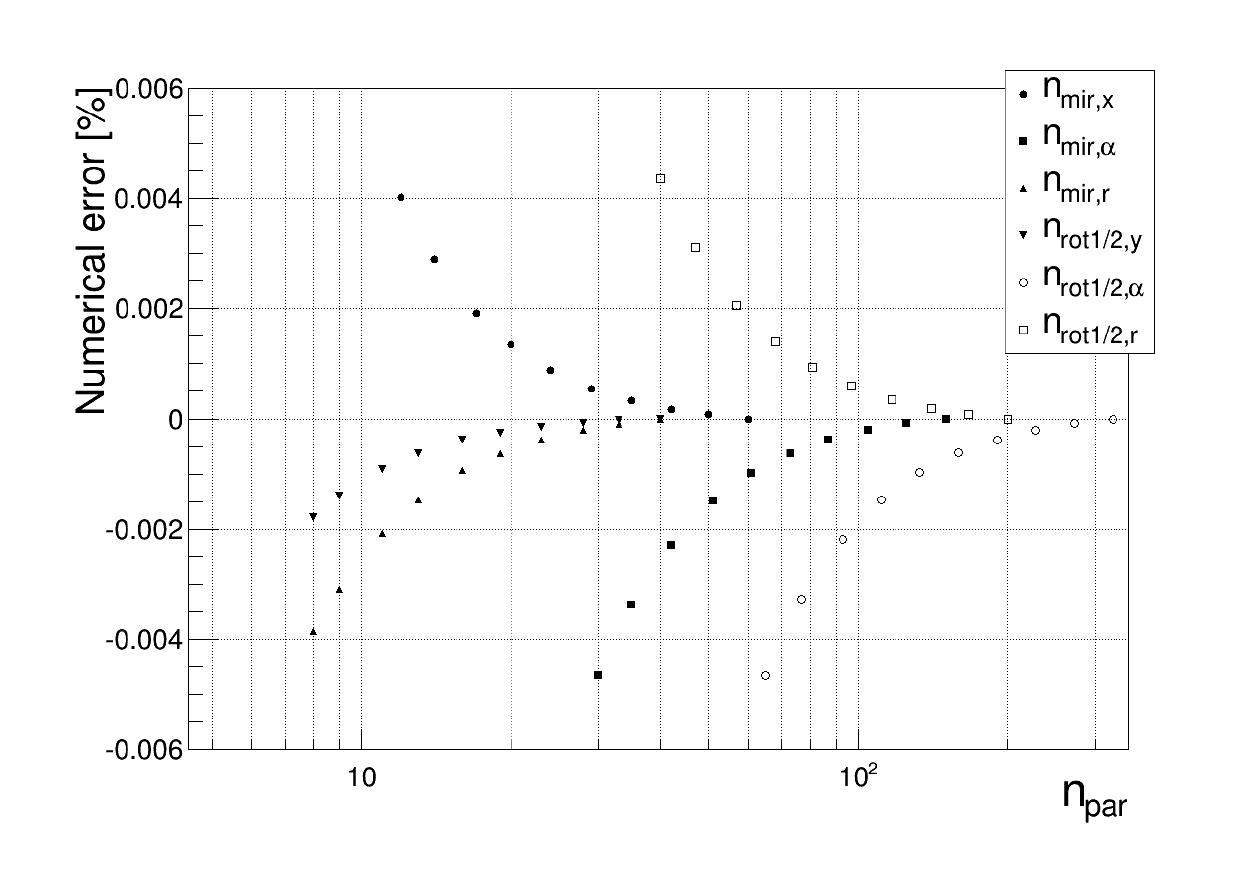} 
	\caption{Numerical error of the NCal results for every parameter of the grid ($n_{par}$) starting with the optimized grid for the mirror $(12,30,8)$ and for a rotor sector $(8,65,40)$.}
	\label{fig:gridsize}
\end{figure}

It is also important to define the grid size adapted to our simulations so that it does not induce a significant numerical error on the results. The grid size used to compute the gravitational force with the Virgo NCal has thus been optimized by doing a convergence test. We call $(n_{mir,x},n_{mir,\alpha},n_{mir,r})$ the mirror grid and $(n_{rot1/2,y},n_{rot1/2,\alpha},n_{rot1/2,r})$ the grid of a rotor sector. Setting the numerical error on the force at $0.005\%$ per grid parameter we found the following optimized grid for the mirror $(12,30,8)$ and for a rotor sector $(8,65,40)$ as shown in figure \ref{fig:gridsize}.
\\ The layout of the Virgo NCal during O3 was made of two rotors respectively at $d_N = 1.2666~$m and $d_F = 1.9466~$m. We first computed the expected amplitude of the mirror motion with FROMAGE for both rotors using simple geometry approximations. The mirror has been set as a full cylinder of radius $175~$mm and thickness $200~$mm and the rotor has been considered as two opposite cylindrical sectors with the parameters given in section \ref{setup}. Then we made refined simulations adding details on the geometry such as the flats, the ears and the anchors which are used to suspend the mirror (see table \ref{tab:mirrorparams}). The holes, the screws, the central aluminum plate and the mechanical fillets for the rotor have also been taken into account according to the rotor plans.
\begin{table}[!h]
\centering
\begin{tabular}{|c|c|c|c|}
\hline
 & Flat & Ear & Anchor \\
\hline 
Length [mm] & 200 & 90 & 39 \\ 
\hline 
Thickness [mm] & - & 10 & 8 \\ 
\hline
Height [mm] & 50 & 15 & 8 \\
\hline
\end{tabular}
\caption{Parameters to define the flats on the mirror and the cuboids to model the ears and anchors.}
\label{tab:mirrorparams}
\end{table}\\
Table \ref{tab:simu} gives the numerical strain of the quadrupole contribution for both rotors using a simple and a more realistic geometry. For the simple geometry case, the strain given by the analytical formula from equation \ref{eq:9} is compared to the numerical results. For the more realistic geometry case, the numerical strain is compared to the simple geometry case numerical strain.\\

\begin{table*}[!h]
\centering
\begin{tabular}{|c|c|c|c|c|}
\hline
     & \multicolumn{2}{|c|}{Simple geometry} & \multicolumn{2}{c|}{More realistic geometry}\\\hline
\hline NCal position & $(d_N, \phi)$ & $(d_F, \phi)$ & $(d_N, \phi)$ & $(d_F, \phi)$\\
\hline 
Numerical strain & $3.35754$e-$18~f_{2rot}^{-2}$ & $6.03812$e-$19~f_{2rot}^{-2}$  & $3.35721$e-$18~f_{2rot}^{-2}$ & $6.03901$e-$19~f_{2rot}^{-2}$\\ 
\hline 
Analytical relative error  & $-0.013\%$ & $-0.004\%$  & - & -\\ 
\hline 
Numerical relative error  & - & -  & $-0.010\%$ & $+0.015\%$ \\ 
\hline 
\end{tabular}
\caption{Comparison of the computed strain between the numerical and analytical models for the NCal near and NCal far located at a distance $d_N = 1.2666~$m (near) and $d_F=1.9466~$m (far) respectively at the real angle $\phi=0.6058~$rad. The relative errors given for the analytical models are computed as $(\frac{analytical}{numerical_{simple}}-1)$ and for the numerical models $(\frac{numerical_{real}}{numerical_{simple}}-1)$. }
\label{tab:simu}
\end{table*}

\noindent
The systematic uncertainty on the computed strain with FROMAGE due to the modeling method could come from three terms:
\begin{itemize}
\item The uncertainty coming from the finite resolution of the grid, which is set to 0.005\% per grid parameter which results in 0.007\% summing all the contributions. This value is an upper limit of the numerical error since it has been computed for the near NCal.
\item The error coming from the modeling method.
This is conservatively taken as the difference between the analytical model and the numerical model (next to last line of table \ref{tab:simu}).
This error is frequency-independent since both models use the same frequency dependence.
\item The error coming from the possible small details of the mirror or rotor geometry that have been ignored. An upper value for this error is the difference between the FROMAGE simulation with the simple geometry and with the more detailed geometry
(last line of table \ref{tab:simu}).
\end{itemize}
Taking the quadratic sum of these three terms, we get the following relative systematic uncertainties from the modeling method on the expected injected strain amplitude: 0.018\% for the near NCal and 0.017\% for the far NCal.  

One of the benefit of the finite element analysis is to make possible the computation of the 
torque induced by the NCal on the mirror, even with a complex geometry.
The contribution of this torque to the longitudinal mirror motion in the gravitational-wave interferometers appears when the main interferometer beam is not centered on the mirrors.
During O3, in Advanced Virgo the position of the beam was controlled to be centered on the mirrors better than $\pm 0.5~$mm. 
Using the maximum values of $\pm0.5~$mm in the $y$ and $z$ directions the contribution to the longitudinal mirror motion is respectively bounded to $\pm 0.05\%$ and $\pm 0.03\%$ for the near and far NCals.
From those values, it is then possible to predict the contribution of the torques for any beam offsets since they are linearly dependent.

Finally, the effect on the NCal-induced displacement coming from the upper stage (\textit{marionette})  to which the mirror is suspended is small.
Indeed, the marionette is located $0.7~$m above the mirror's center which gives a larger marionette to rotor distance than the mirror to rotor distance.
The induced amplitude of the marionette is thus almost a factor 2 smaller than the mirror's amplitude. Then, this displacement is filtered by the mirror's suspension which can be modelled by a simple pendulum transfer function.
For frequencies greater than the resonant frequency of the pendulum $f_0 = 0.6~$Hz, the effect of the marionette on the mirror is reduced by a factor $f_0^2/f^2$. At $10~$Hz, the effect of the marionette is equal to $1.8$e-$3$ of the direct mirror's displacement and goes quickly to lower values as the frequency increases. Therefore, we neglect it for this study but this will have to be taken into account for future measurements.

\section{Data set used and raw calibration check } \label{dataset}
Several data set were collected during O3, that allow us to adjust the NCal calibration procedure.
We describe here the last and longer test, six hours of data collected on March 24, 2020 with the Virgo interferometer running at its nominal sensitivity (around 55 Mpc),
just before the premature end of the run triggered by the Covid-19 pandemic.
During this calibration period, both NCals were used, scanning their frequency band.
Their rotation speed were set to close values for part of the measurements in order to remove frequency-dependent effects when extracting the mirror to NCal distances, as it will be described later.

\begin{figure} [h]
\centerline{\includegraphics[width=0.9\textwidth]{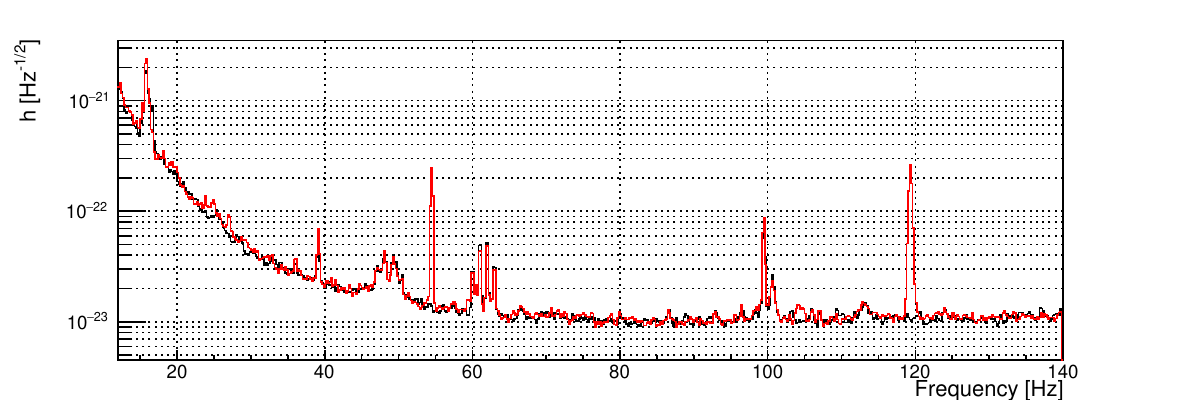}}
\caption{red trace: $h(t)$ spectrum when the far NCal was injecting a line at 55 Hz, and the near 
NCal a line at 119 Hz. black trace: spectrum when the NCal was switched off. The FFTs are averaged over 200 seconds.}
\label{fig:hoftSpectrum}
\end{figure}

Figure \ref{fig:hoftSpectrum} presents the $h(t)$ spectrum observed with the near NCal at its maximum rotation speed injecting a line at $119~$Hz and the far NCal injecting a line at $55~$Hz, compared to the spectrum taken an hour later, when the NCals were switched off.
The differences are marginal, besides the two NCal lines at 55 and 119 Hz.
The NCal lines are not perfectly monochromatic, especially at 119 Hz, due to fluctuations of the NCal rotation since the NCal motor was a standalone device with just a simple commercial driver.

It is worth noting that despite this time was tagged as calibration time, and therefore not used for regular analysis, the NCals operation is not very invasive with usually just two additional stable calibration lines that could be subtracted. Therefore, if an exceptional event would have been detected during this time, the data could probably have been used.

The NCal motor used during O3 was a standalone system, or in other words, the NCal rotation was independent of the Virgo timing system. 
The phase and rotation speed was therefore extracted from the data, using the pulses observed by a photodiode each time the rotor hole is in front of the LED. This signal was recorded by the Virgo data acquisition.
The time of each individual pulse was computed by taking the average value of the time of the rising and falling fronts. 
Each of these times is computed by doing a linear interpolation of the time of the samples around the crossing of 50 \% of the pulse height, going up or down.
This method gives the central value of the pulse and is independent of the rotor speed.
Once the pulse times are known, the rotation period is just the time difference between two pulses and the instantaneous frequency the invert of this period. 

The NCal lines amplitude is extracted by doing the transfer function between  the $h(t)$ channel and a signal describing the effective NCal rotation to be immune to small variations of the rotor rotation.
For this purpose, a set of sine-waves of unit amplitude is built with the same sampling rate as the $h(t)$ channel. 
Their frequencies are N times the rotor instantaneous frequency (typical value: N=2).
The NCal $h(t)$ amplitude for the harmonic N of the rotor is then computed by reading, at the harmonic frequency, the transfer function value between the reconstructed $h(t)$ and the corresponding sine wave. 
This transfer function is computed over 20 seconds, meaning that a new value of the NCal $h(t)$ recovered amplitude is available with the same period.
These 20 seconds have been selected to keep a good signal-to-noise ratio without losing too much signal due to the small variations of the rotation speed.
The phase is extracted in the same way.
Given the position of the hole in the rotor as well as the LED/photodiode position, when the two massive sectors are vertical, the phase is zero. 

\begin{figure} [h]
\centerline{\includegraphics[width=1.05\textwidth]{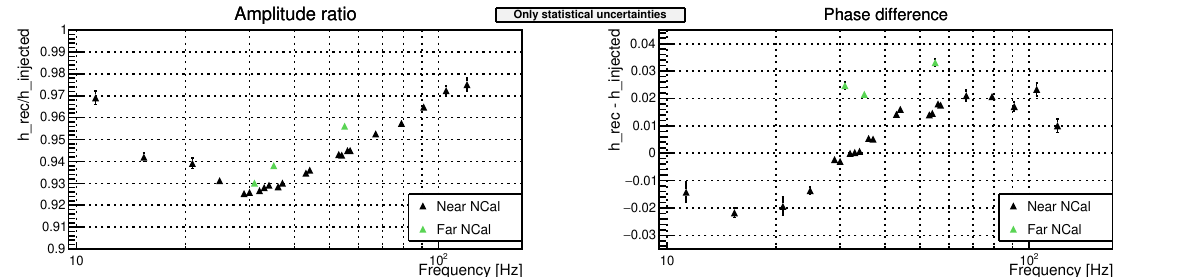}}
\caption{Comparing the recovered NCal lines ($h_{rec}$) amplitude and phase to the expected values ($h_{injected}$) computed with FROMAGE and using the expected NCal to mirror relative positions.}\label{fig:rawCheck}
\end{figure}

Once the reconstructed NCal line amplitudes are computed and averaged over some time, they can be compared to the expected values or "injected" signal, computed using the FROMAGE model.
The result is presented in figure \ref{fig:rawCheck}, where only statistical uncertainties are shown.
Deviations from the expected values of 1 for the ratio, and 0 for the phase difference are observed and will be addressed later in the paper. 
But more interesting at this stage of the study is to notice the differences between the near and far NCals, that are supposed to be identical, coming from the same production batch.
Such difference could easily be explained by slight deviations  of the NCal position relative to the true mirror location and will be discussed in  the following sections.

\section{Extracting the NCal mirror distance } \label{distance}
The main discrepancy between the near and far NCal amplitude points may be explained by a common offset $d_0$ in the distance of both NCals to the mirror. Since the distance between both NCals is fixed and known, $d_{NF} = 680~$mm, it is possible to extract a measured distance by comparing the NCal amplitudes and correct the predicted NCal signal accordingly.

\begin{figure}[h!] 
  \centering
  \includegraphics[trim={0.5cm 1cm 0cm 0cm},clip,scale=0.55]{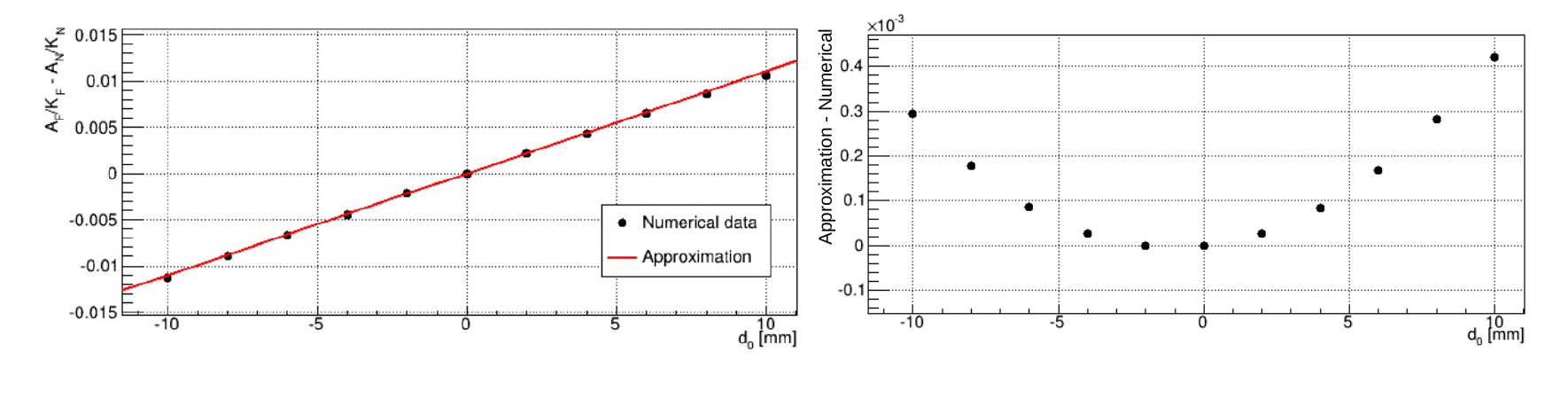}  
    \caption{Left: far minus near NCal relative amplitudes as a function of the distance offset for the linear approximation of equation \ref{eq:diffratio} and the numerical results computed with FROMAGE. Right: differences between both models.}
	\label{numapproxdiff}
\end{figure}

Let's call $A_i$ the amplitude of the mirror motion induced by the NCal$\_i$ with $i=\lbrace N,F \rbrace$ at the distance $d_i+d_0$. The amplitude of the time dependent gravitational force at twice the rotor frequency can be expressed as:
\begin{equation}
A_i = C_i(d_i+d_0)^{-4}
\label{rawamp}
\end{equation}
with $C_i$ the gravitational coupling factor. From equation \ref{rawamp} and assuming $d_0 \ll d_i$ we can write:
\begin{equation}
A_i \approx K_i(1-4\frac{d_0}{d_i})
\label{newamp}
\end{equation}
with $K_i$ the amplitude of the mirror motion computed with a rotor at the estimated distance $d_i$ from the mirror. The difference of ratios between the real amplitude $A$ and the estimated amplitude $K$ of both NCals is:
\begin{equation}
\frac{A_F}{K_F}-\frac{A_N}{K_N} \approx 4d_0\frac{d_F-d_N}{d_F \cdot d_N} \approx 1.103~d_0
\label{eq:diffratio}
\end{equation}
One can check that this approximation holds for small offsets $d_0$, typically smaller than $1~$cm, using FROMAGE. Figure \ref{numapproxdiff} shows the comparison between the numerical results and the approximation of equation~\ref{eq:diffratio}. The linear approximation is in agreement with the numerical results within better than $0.05\%$ for any offset $d_0\leq 1~$cm.
\\
Table \ref{tab:extractdistance} gathers the far and near NCals $h_{rec}/h_{inj}$ ratios used to compute distance offset $d_0$.
For each measurement with the far NCal, a couple of measurement at frequencies
that bracket the far NCal frequency were performed with the near NCal, 
to minimize some possible frequency dependent response.
Then the average value of the two near NCal measurements at the "low" and "high" frequencies is made 
and formula \ref{eq:diffratio} is used to compute 
the corresponding $d_0$ values reported in this table.

\begin{table}[!h]
\centering
\begin{tabular}{|c|c|c|c|c|c|c|c|}
\hline
  \multicolumn{2}{|c|}{Far NCal} & \multicolumn{5}{c|}{Near NCal} & \\
\hline Freq. [Hz] & $h_{rec}/h_{inj}$ & f. low [Hz] & $h_{rec}/h_{inj}$ & f. high [Hz] & $h_{rec}/h_{inj}$ & $<h_{rec}/h_{inj}>$ & $d_0$ [mm]\\
\hline 31.0 & 0.9301 & 30.0  & 0.9258 & 32.0 & 0.9267 & 0.92625 & 3.5\\ 
\hline 35.0  & 0.9381 & 34.0  & 0.9292 & 36.0 & 0.9285 & 0.92885 & 8.4\\ 
\hline 55.0  & 0.9561 & 54.0  & 0.9429 & 56.0 & 0.9449 & 0.9439 & 11.1\\ 
\hline 
\end{tabular}
\caption{Values of the $h_{rec}/h_{inj}$ ratios for the far and near NCals at different frequencies and estimated distance offset $d_0$.}
\label{tab:extractdistance}
\end{table}

\noindent
The three $d_0$ values report a systematic offset, with an average value of 7.6 mm computed with the exact $d_0$ values (the table reports rounded values).
Removing part of the measurement time does not change significantly each of the $d_0$ values.
The statistical uncertainty on each $d_0$, derived  from the $h_{rec}/h_{inj}$ errors, is
of the order of 1 mm, well below the observed dispersion of the three values.
The three $d_0$ values are increasing with frequency but this is not expected for a distance offset.
In a normal situation, this would call for extra measurements, that were expected to be made during the end of run period.
But due to the Covid-19 pandemic, 
the end of run was anticipated and the interferometer was shutdown a few days after the measurements reported in this paper.

Therefore, the measurement uncertainty associated to $d_0$ is set to the maximum difference between the mean value and the farthest value: 4.1 mm (here again, the difference has been computed with the exact $d_0$ values and not the rounded ones from the table). 
 
This method of extracting $d_0$ assumes identical NCals. 
This could have been checked by swapping the two NCals and repeating the $d_0$ measurement procedure.
But again, it was not possible due to the Covid-19 pandemic.
Therefore we use the  $0.53\%$ amplitude uncertainty from the rotor geometry as  
the maximum difference between the signals produced by the two rotors and derive the corresponding error on the measured distance: 4.8 mm.

Finally, we do the quadratic sum of both uncertainties and obtain: $d_0 = 7.6 \pm 6.4~$mm

\section{Extracting the NCal offset with respect to the plane of the interferometer} \label{offset}
\begin{figure}[h!] 
  \centering
  \includegraphics[trim={1cm 1cm 1cm 1cm},clip,scale=0.64]{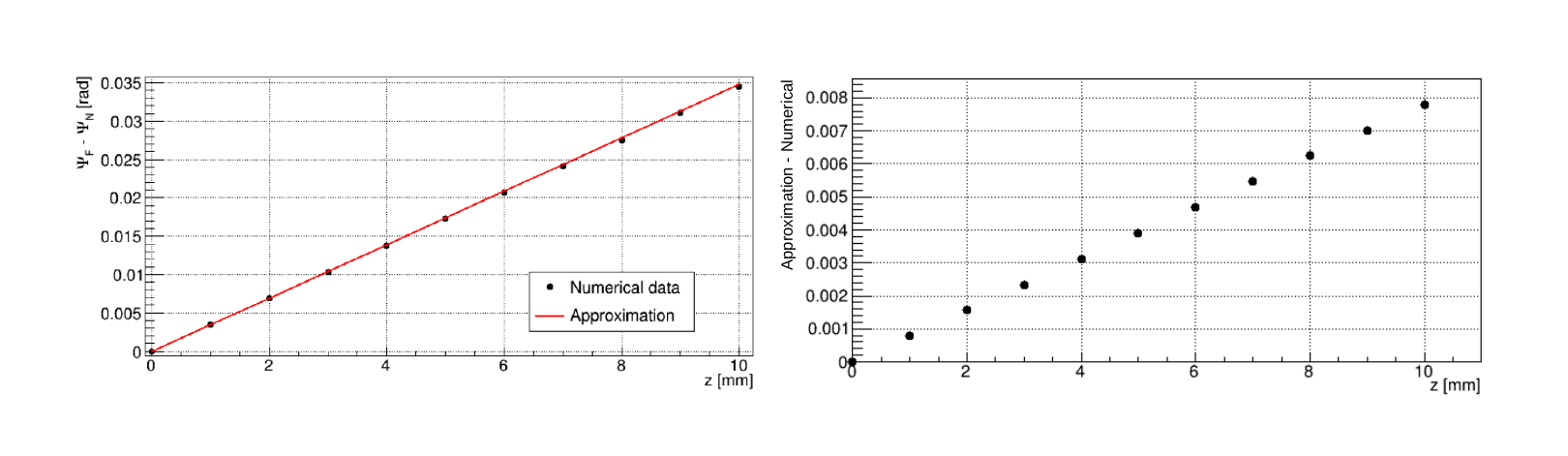}  
    \caption{Left: far minus near NCal phase as a function of the vertical offset for the linear approximation of equation \ref{eq:phaseoffset} and the numerical results computed with FROMAGE. Right: differences between both models.}
	\label{fig:offset}
\end{figure}

The phase difference between the near and far NCal points is mainly driven by the offset $z$ of the rotor with respect to the plane of the interferometer. 
From the calculations in appendix \ref{app:analytical}, the real analytical phase offset for the NCal$\_i$ at the first order in $d^{-1}$ can be approximated by:
\begin{equation}
\Psi_{i} \approx -\frac{8}{3}\frac{z}{d_i}
\label{eq:phaseapprox}
\end{equation}
The rotors have been set on Advanced Virgo such that they spin in opposite directions. 
They thus have an opposite sign on their phase and it is possible to extract the offset using this phase difference. 
Let's call $\xi_i$ the phase of the mirror motion with the NCal$\_i$ whose rotor is in the plane of the interferometer $(z=0)$.  In this case, $\xi_F = \xi_N$. Assuming the offset $z$ to be the same for both NCals the difference between the phase points can be computed as:
\begin{align}
(\Psi_F-\xi_F) - (\Psi_N-\xi_N) & \approx \pm\frac{8}{3}(\frac{z}{d_F}+\frac{z}{d_N}) \nonumber \\
& \approx \pm\frac{8}{3}\frac{d_N + d_F}{d_F d_N}z \nonumber \\
& \approx \pm3.475~z
\label{eq:phaseoffset}
\end{align}
Assuming $\Psi_{N} <0$ and $\Psi_{F} > 0$ without any loss of generality, we can verify that this approximation holds using FROMAGE for small offsets $z$ with typical values smaller than $1~$cm. Figure \ref{fig:offset} shows the comparison between the numerical results and the approximation of equation \ref{eq:phaseoffset}. The linear approximation differs from the numerical results by a quasi-constant value of $0.81\%$ on the slope for any offset $z \leq 1~$cm.  
This difference mainly comes from the fact that we did not take into account the second order terms in $d^{-2}$ in equation \ref{eq:phaseapprox}. 
\\
We can thus normalize the phase for a better approximation by redefining $\Psi_i$ as:
\begin{equation}
\Psi_i \approx - \frac{8}{3}\frac{z}{d_i}(1-0.0081)
\label{eq:phaseoffsetaprox}
\end{equation}
Table \ref{tab:extractoffset} gives the values of the phase at different frequencies for the near and far NCals with the estimated offset $z$. 
As for the distance measurements, we compute the average value of these offsets and report it in the first line of table \ref{tab:erroroffset}. 
This line presents also the phase shifts computed using equation \ref{eq:phaseoffsetaprox}.
The quoted $z$ uncertainty on the second line, that is also translated to phase uncertainties, is the linear sum of two terms.
The first one, 1.0 mm, is the maximum difference between the mean value and the farthest $z$ value. 
The second term is from the mechanical uncertainty in the relative phase origin, 0.5 mrad, 
which translates to twice this value at the NCal signal frequency 
and finally to a $z$ systematic uncertainty of 0.3 mm.
Finally, the last line of the table is the overall systematic phase uncertainty.
This is the quadratic sum of the second line numbers plus 2 mrad. These 2 mrad come from the 1 mrad mechanical absolute phase origin uncertainty,
which again should be doubled at the NCal signal frequency.
\begin{table}[!h]
\centering
\scalebox{0.9}{
\begin{tabular}{|c|c|c|c|c|c|c|c|}
\hline
  \multicolumn{2}{|c|}{Far NCal} & \multicolumn{5}{c|}{Near NCal} & \\
\hline Freq. [Hz] & $h_{rec} -h_{inj}$ & f. low [Hz] & $h_{rec}-h_{inj}$ & f. high [Hz] & $h_{rec}-h_{inj}$ & $<h_{rec}-h_{inj}>$ & $z$ [mm]\\
\hline 31.0 & 24.7 & 30.0  & 3.0 & 32.0 & 0 & 1.5 & 6.6\\ 
\hline 35.0  & 21.5 & 34.0  & 0.7 & 36.0 & 5.4 & 3.05 & 5.3\\ 
\hline 55.0  & 33.1 & 54.0  & 14.6 & 56.0 & 18.0 & 16.3 & 4.8\\ 
\hline 
\end{tabular}}
\caption{Values of the phase difference $h_{rec} -h_{inj}$ (in mrad) with the far and near NCals at different frequencies and estimated offset $z$. For the near NCal, the average of the phase measured around the frequency of the far NCal is done to extract the phase at the far NCal frequency.}
\label{tab:extractoffset}
\end{table}
\begin{table}[!h]
\centering
\begin{tabular}{|c|c|c|c|}
\hline
 & $z$ [mm] & $h_{rec} -h_{inj}$ near [mrad] & $h_{rec}-h_{inj}$ far [mrad]\\
\hline
$z$ and phase difference offsets & 5.6 & 11.6 & -7.6 \\
\hline Uncertainties from phase difference & 1.3 & 2.7 & 1.8 \\ 
\hline
 \multicolumn{2}{|c|}{Overall phase uncertainties} & 3.4 & 2.7 \\ 
\hline 

\end{tabular}
\caption{Value of the measured $z$ and its systematic uncertainty.  The change in the phase difference $h_{rec} - h_{inj}$ for both NCals is also reported with the associated systematic uncertainties as well as the overall uncertainties.}
\label{tab:erroroffset}
\end{table}

\section{Checking the $h(t)$ reconstruction with the NCal and PCal } \label{check}
\begin{figure} [h]
\centerline{\includegraphics[width=1.1\textwidth]{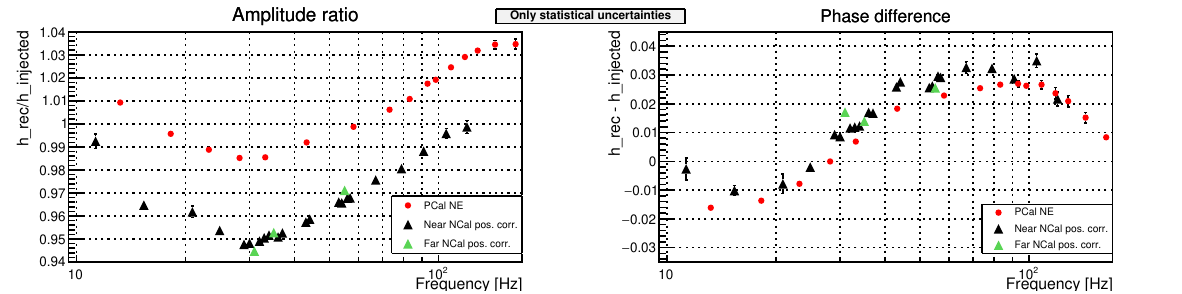}}
\caption{Comparing recovered ($h_{rec}$) to injected ($h_{injected}$) calibration lines for PCal and for NCal at the adjusted position.}\label{fig:finalCheck}
\end{figure}

Using the measured distance and vertical offset for both NCals given in previous sections, 
we recomputed the new expected injected amplitudes and phases of the mirror motion with FROMAGE.
These new values have thus been used to correct the data points of the $h_{rec}$ to $h_{inj}$  comparison and are shown in figure \ref{fig:finalCheck}. 
The error bars reported on the figures are just statistical uncertainties.

Several contributions to the systematic uncertainties on the expected amplitude were discussed in previous sections. 
The NCal to mirror angle uncertainty coming from the NCal positioning is expected to be small.  
But since there is a large measured offset in the NCal to mirror distance, we assumed the same offset as maximum
possible lateral shift of the NCals and derived the corresponding angles for the two NCals.
Table \ref {table:uncertainties} summarizes them, and presents the overall values.

The systematic uncertainty of the NCal phase is expected to primarily come from the uncertainty on the NCal signals readout timing of 10 $\upmu$s, 
which translates to a phase uncertainty of 6 mrad at 100 Hz to be added with the uncertainty given in table \ref{tab:erroroffset}.

The NCal points are compared to the reference method of calibration for Advanced Virgo during O3 using photon calibrators.
The principle of PCals is to induce a known mirror motion using radiation pressure of auxiliary lasers. 
The systematic uncertainty associated to the PCal points used in figure \ref{fig:finalCheck} is $1.4\%$. These measurements were performed with the PCal acting on the North End (NE) test mass of Virgo which is the same test mass used for the NCal measurements. 

The frequency dependent shape of the NCal amplitude points is similar to the PCal ones but a difference of about $3\%$ is observed between both techniques. 
This difference, is not incompatible with the sum of the systematic uncertainties of the two methods, that are expected to be frequency independent.
On the phase, the NCal and PCal points are in good agreement, both regarding the shape and the absolute values. 

\begin{table}[h!]
\begin{center}
\begin{tabular}{|c|c|c|c|c|}
\hline
Parameter & uncertainty     & formula & $h_{rec}/h_{inj}$ near [$\%$]  & $h_{rec}/h_{inj}$ far [$\%$] \\
\hline

NCal to mirror distance $d$   & 6.4 mm & $4 \delta d/d$ & 2.02 &1.31\\
NCal to mirror angle $\Phi$ & 5.0/3.3 mrad& $\delta\Phi \sin\Phi$ & 0.28 & 0.19\\
NCal vertical position $z$ &  1.3 mm & ${5/2(z/d)^2}$  & 0.03 &  0.01 \\
\hline
Rotor geometry &\multicolumn{2}{|c|}{see table \ref{table:rotor-uncertainties}}& 0.53 & 0.53\\
\hline
Modeling method&\multicolumn{2}{|c|}{see end of section \ref{fromage}}& 0.018& 0.017\\
\hline
Mirror torque from NCal &\multicolumn{2}{|c|}{see end of section \ref{fromage}}& 0.05& 0.03\\
\hline
Total &\multicolumn{2}{|c|}{quadratic sum}  & 2.1& 1.4\\
\hline
\end{tabular}
\caption{Uncertainties on calibration signal amplitude}
\label{table:uncertainties}
\end{center}
\end{table}

\section{Conclusion and outlook} \label{conclusions}
A new version of NCal was successfully tested on the Virgo detector during O3. 
It was possible to check the $h(t)$ reconstruction up to 120 Hz, and to confirm the shape observed with the reference method that uses a photon calibrator.
This result indicates that the frequency-dependent bias is coming from the reconstruction algorithm of $h(t)$.  
A difference of about 3\% was observed, that is of the order of the sum of the systematic uncertainties of both methods.

The anticipated end of the O3 data taking due to the Covid-19 pandemic prevented us to perform all the needed measurements, 
like determining the relative strain induced by the two NCals, 
the dominant term in our amplitude systematic uncertainty.
Nevertheless, the achieved amplitude systematic uncertainty for the far NCal was the same as for the PCal, 
confirming the potential of the NCal technique.

We are planning to build additional and slightly upgraded NCals for the upcoming O4 data run, scheduled to start in 2022.
The aim is to simplify the current rotor geometry to reduce the corresponding systematic uncertainty.
We also plan to optimize the rotor dimensions to increase the NCal-induced signal.
Having more of them, especially on both sides of the mirror, 
will allow us to have a better measurement of the mirror to NCal distance and bring the systematic uncertainty well below 1\%.
With such a level of uncertainty on the NCal-induced displacement it would be interesting to cross-calibrate the PCal at low frequency with the NCal and then extend the calibration of Virgo with the PCal at higher frequencies. 

\section*{Acknowledgements}
We are indebted to the Virgo Collaboration for allowing us to use the data
collected during the tests reported here, and are grateful for support
provided by the Virgo Collaboration and the European Gravitational
Observatory during those tests. 
We thank our colleagues in the Virgo Collaboration and in the LIGO Scientific Collaboration for useful discussions. 
We thank the technical staff at LAPP for their help in building the O3 NCals, 
especially B. Lieunard for the mechanical design and installation, L. Journet for the assembly, 
G. Deleglise for the metrology and S. Petit for the motor control. 
We also thanks D. Dangelser from IPHC for improving the NCal support system.

\section*{References}
\let\clearpage\relax

\newpage
\section*{Appendix} \label{app1}
\renewcommand{\thesubsection}{\Alph{subsection}}
\subsection{Analytical calculation of the Virgo NCal-induced mirror motion}
\label{app:analytical}

We provide here a detailed calculation of the Virgo NCal-induced mirror amplitude in an off-axis configuration and with an offset $z$ with respect to the plane of the interferometer. The coordinates of a point of the rotor masses $(+,-)$ are:
$$\begin{pmatrix}d\cos(\phi)\pm r\cos(\theta+\psi)\cos(\phi) - b'\sin(\phi) \\ d\sin(\phi)\pm r\cos(\theta+\psi)\sin(\phi) + b'\cos(\phi) \\ z \pm r\sin(\theta+\psi) \end{pmatrix}$$
and the coordinates of a point of the mirror are:
$$(x';r'\cos(\beta);r'\sin(\beta))$$
The gravitational force induced by a point of a rotor sector on a point of the mirror along the $x$ axis is given by:
\begin{equation}
F_{\pm,x} = \frac{GmM}{D^{3/2}}(d\cos(\phi)\pm r \cos(\theta+\psi)\cos(\phi)-b'\sin(\phi)-x')
\label{eq:app1}
\end{equation}
where $G$ is the gravitational constant, $m$ is the mass of the masses $(+,-)$, $M$ is the mass of the mirror, $d$ is the distance from the center of the rotor to the center of the mirror in the plane of the interferometer and $D$ is the distance between a point of a rotor sector and a point of the mirror:
\begin{align}
D & = \Big[(d\cos(\phi)\pm r \cos(\theta+\psi)\cos(\phi)-b'\sin(\phi)-x')^2 \nonumber \\
&~~~~~+(d\sin(\phi)\pm r\cos(\theta+\psi)\sin(\phi) + b'\cos(\phi)-r'\cos(\beta))^2 \nonumber \\
&~~~~~+(z \pm r\sin(\theta+\psi)-r'\sin(\beta))^2 \Big]^{1/2}
\label{eq:app2}
\end{align}
We define the small quantities $\epsilon = \frac{r}{d}$, $\epsilon' = \frac{r'}{d}$, $\epsilon'' = \frac{x'}{d}$, $\epsilon''' = \frac{b'}{d}$ and $z' = \frac{z}{d}$, the longitudinal force then becomes:
\begin{equation}
F_{\pm,x} = \frac{GmM}{d^2}(\cos(\phi)\pm\epsilon \cos(\theta+\psi)\cos(\phi)-\epsilon'''\sin(\phi)-\epsilon'')\Big[1+X_{\pm}\Big]^{-3/2}
\label{eq:app3}
\end{equation}
where:
\begin{align}
X_{\pm} & = \epsilon^2+\epsilon'^2+\epsilon''^2+\epsilon'''^2\pm2\epsilon\cos(\theta+\psi)-2\epsilon''\cos(\phi)(1\pm\epsilon\cos(\theta+\psi)) \nonumber \\
&~~-2\epsilon'\sin(\phi)\cos(\beta)(1\pm\epsilon\cos(\theta+\psi))\mp 2\epsilon\epsilon'\sin(\beta)\sin(\theta+\psi)) \nonumber \\
&~~+2\epsilon'''(\epsilon''\sin(\phi)-\epsilon'\cos(\phi)\cos(\beta))\pm 2\epsilon z'\sin(\theta+\psi) - 2\epsilon' z' \sin(\beta)
\label{eq:app4}
\end{align}
The expression from equation \ref{eq:app3} can be expanded at the fourth order in $X_{\pm}$ using:
\begin{equation}
 (1+X_{\pm})^{-3/2} \approx\Big(1-\frac{3}{2}X_{\pm}+\frac{15}{8}X_{\pm}^2-\frac{35}{16}X_{\pm}^3+\frac{315}{128}X_{\pm}^4\Big)
\label{eq:app5}
\end{equation}
Let's define:
\begin{equation}
Y_{1} = (\cos(\phi)+\epsilon \cos(\theta+\psi)\cos(\phi)-\epsilon'''\sin(\phi)-\epsilon'')X_{+}+(\cos(\phi)-\epsilon \cos(\theta+\psi)\cos(\phi)-\epsilon'''\sin(\phi)-\epsilon'')X_{-}
\label{eq:app6}
\end{equation}
\begin{equation}
Y_{2} = (\cos(\phi)+\epsilon \cos(\theta+\psi)\cos(\phi)-\epsilon'''\sin(\phi)-\epsilon'')X_{+}^2+(\cos(\phi)-\epsilon \cos(\theta+\psi)\cos(\phi)-\epsilon'''\sin(\phi)-\epsilon'')X_{-}^2
\label{eq:app7}
\end{equation}
\begin{equation}
Y_{3} = (\cos(\phi)+\epsilon \cos(\theta+\psi)\cos(\phi)-\epsilon'''\sin(\phi)-\epsilon'')X_{+}^3+(\cos(\phi)-\epsilon \cos(\theta+\psi)\cos(\phi)-\epsilon'''\sin(\phi)-\epsilon'')X_{-}^3
\label{eq:app8}
\end{equation}
\begin{equation}
Y_{4} = (\cos(\phi)+\epsilon \cos(\theta+\psi)\cos(\phi)-\epsilon'''\sin(\phi)-\epsilon'')X_{+}^4+(\cos(\phi)-\epsilon \cos(\theta+\psi)\cos(\phi)-\epsilon'''\sin(\phi)-\epsilon'')X_{-}^4
\label{eq:app9}
\end{equation}
Keeping only the time dependent terms at twice the rotor frequency up to the fourth order, omitting the odd terms in $\cos(\beta)$, $\sin(\beta)$, $\epsilon''$ and $\epsilon'''$ since they further cancel when integrating over the mirror and the rotor and defining $\Theta = \theta + \psi$ we get:
\begin{align}
Y_{1} & \approx \cos(\phi)\Big[4\epsilon^2\cos^2(\Theta) + 4 \epsilon^2 z' \cos(\Theta) \sin(\Theta) \Big]
\label{eq:app10}
\end{align}
\begin{align}
Y_{2} & \approx \cos(\phi)\Big[8\epsilon^2\cos^2(\Theta) + 8\epsilon^4\cos^2(\Theta) + 8\epsilon^2 \epsilon'^2 \cos^2(\Theta) + 24 \epsilon^2 \epsilon''^2 \cos^2(\Theta) + 8\epsilon^2 \epsilon'''^2 \cos^2(\Theta) \nonumber \\
&~~~~~~~~~~~~~+ 24 \epsilon^2 \epsilon''^2 \cos^2(\phi) \cos^2(\Theta) + 16 \epsilon^2 z' \cos(\Theta) \sin(\Theta) + 24 \epsilon^2 \epsilon'^2 \sin^2(\phi) \cos^2(\beta) \cos^2(\Theta) \nonumber \\
&~~~~~~~~~~~~~+8\epsilon^2 \epsilon'^2 \sin^2(\beta) \sin^2(\Theta) \Big]
\label{eq:app11}
\end{align}
\begin{align}
Y_{3} & \approx \cos(\phi)\Big[24\epsilon^4\cos^2(\Theta) + 24\epsilon^2 \epsilon'^2 \cos^2(\Theta) + 72 \epsilon^2 \epsilon''^2 \cos^2(\Theta) + 24\epsilon^2 \epsilon'''^2 \cos^2(\Theta)+ 16\epsilon^4\cos^4(\Theta) \nonumber \\
&~~~~~~~~~~~~~+ 144 \epsilon^2 \epsilon''^2 \cos^2(\phi) \cos^2(\Theta) + 24 \epsilon^2 z'^2 \cos^2(\Theta)  + 144 \epsilon^2 \epsilon'^2 \sin^2(\phi)\cos^2(\beta)\cos^2(\Theta) \Big]
\label{eq:app12}
\end{align}
\begin{equation}
Y_{4} \approx \cos(\phi)\Big[32\epsilon^4\cos^4(\Theta) + 192\epsilon^2 \epsilon'^2 \sin^2(\phi) \cos^2(\beta) \cos^2(\Theta) + 192 \epsilon^2 \epsilon''^2 \cos^2(\phi) \cos^2(\Theta)\Big]
\label{eq:app13}
\end{equation}
The total time dependent longitudinal force can then be expressed as:
\begin{align}
F_{x} & \approx \frac{GmM}{d^2}\Big(-\frac{3}{2}Y_1+\frac{15}{8}Y_2-\frac{35}{16}Y_3+\frac{315}{128}Y_4 \Big) \nonumber \\
& \approx \frac{GmM}{d^2} \cos(\phi) \Big[\Big(\frac{9}{2}\epsilon^2 + \frac{25}{8} \epsilon^4 + \Big(\frac{405}{4}\sin^2(\phi)\cos^2(\beta)-\frac{15}{2}\sin^2(\beta)-\frac{75}{4}\Big)\epsilon^2 \epsilon'^2 \nonumber \\
&~~~~~~~~~~~~~~~~~~~~~~~~~+\Big(\frac{405}{4}\cos^2(\phi)-\frac{225}{4}\Big)\epsilon^2 \epsilon''^2 - \frac{75}{4} \epsilon^2 \epsilon'''^2 - \frac{105}{4} \epsilon^2 z'^2 \Big)\cos(2\Theta)+ 12 \epsilon^2 z' \sin(2\Theta) \Big] \nonumber \\
& \approx \frac{9GmMr^2}{2d^4} \cos(\phi) \Big[\Big(1 + \frac{25}{36} \epsilon^2 + \Big(\frac{45}{2}\sin^2(\phi)\cos^2(\beta)-\frac{5}{3}\sin^2(\beta)-\frac{25}{6}\Big)\epsilon'^2 \nonumber \\
&~~~~~~~~~~~~~~~~~~~~~~~~~~~~~+\Big(\frac{45}{2}\cos^2(\phi)-\frac{25}{2}\Big)\epsilon''^2 - \frac{25}{6} \epsilon'''^2 - \frac{35}{6} z'^2 \Big)\cos(2\Theta)+ \frac{8}{3} z' \sin(2\Theta) \Big]
\label{eq:app14}
\end{align}
The mass of a rotor sector and of the mirror are:
\begin{equation}
m=\rho_{rot}\int_{r_{min}}^{r_{max}} \int_{-\frac{\alpha}{2}}^{\frac{\alpha}{2}} \int_{-\frac{b}{2}}^{\frac{b}{2}} r~dr~d\psi~db'
\label{eq:app15}
\end{equation}
\begin{equation}
M=\rho_{mir}\int_{0}^{r_{mir}} \int_{0}^{2\pi} \int_{-\frac{x_{mir}}{2}}^{\frac{x_{mir}}{2}} r'~dr'~d\beta~dx'
\label{eq:app16}
\end{equation}
with $\rho_{rot}$ and $\rho_{mir}$ respectively the density of the rotor material and the mirror material.\\
Integrating equation \ref{eq:app14} using equations \ref{eq:app15} and \ref{eq:app16} the NCal longitudinal force becomes:
\begin{align}
F_{x} & \approx \frac{9G}{8d^4}\rho_{rot}b(r_{max}^4-r_{min}^4)\sin(\alpha) \rho_{mir} \pi r_{mir}^2 x_{mir} \cos(\phi) \Big[\Big(1 + \frac{25}{54d^2} \frac{r_{max}^6-r_{min}^6}{r_{max}^4-r_{min}^4} \nonumber \\
&~~~~~~~~~~~~~~~~~~~~~~~~~~~~~~~~~~~~~~~~~~~~~~~~~~~~~~~~~~~~~~~~~~~~~~~~~+ \Big(\frac{45}{8}\sin^2(\phi)-\frac{5}{2}\Big)\Big(\frac{r_{mir}}{d}\Big)^2 \nonumber \\
&~~~~~~~~~~~~~~~~~~~~~~~~~~~~~~~~~~~~~~~~~~~~~~~~~~~~~~~~~~~~~~~~~~~~~~~~~+\Big(\frac{15}{8}\cos^2(\phi)-\frac{25}{24}\Big)\Big(\frac{x_{mir}}{d}\Big)^2 \nonumber \\
&~~~~~~~~~~~~~~~~~~~~~~~~~~~~~~~~~~~~~~~~~~~~~~~~~~~~~~~~~~~~~~~~~~~~~~~~~- \frac{25}{72} \Big(\frac{b}{d}\Big)^2 - \frac{35}{6} \Big(\frac{z}{d}\Big)^2 \Big)\cos(2\theta) \nonumber \\
&~~~~~~~~~~~~~~~~~~~~~~~~~~~~~~~~~~~~~~~~~~~~~~~~~~~~~~~~~~~~~~~~~~~~~~~~~+ \frac{8}{3} \frac{z}{d} \sin(2\theta) \Big] \nonumber \\
& \approx \frac{9G}{8d^4}\rho_{rot}b(r_{max}^4-r_{min}^4)\sin(\alpha) \rho_{mir} \pi r_{mir}^2 x_{mir} \cos(\phi)\Big[A\cos(2\theta) + B\sin(2\theta) \Big]
\label{eq:app17}
\end{align}
where:
\begin{equation}
A = 1 + \frac{25}{54d^2} \frac{r_{max}^6-r_{min}^6}{r_{max}^4-r_{min}^4} + \Big(\frac{45}{8}\sin^2(\phi)-\frac{5}{2}\Big)\Big(\frac{r_{mir}}{d}\Big)^2 +\Big(\frac{15}{8}\cos^2(\phi)-\frac{25}{24}\Big)\Big(\frac{x_{mir}}{d}\Big)^2 - \frac{25}{72} \Big(\frac{b}{d}\Big)^2 - \frac{35}{6} \Big(\frac{z}{d}\Big)^2
\label{eq:app18}
\end{equation}
\begin{equation}
B = \frac{8}{3} \frac{z}{d}
\label{eq:app19}
\end{equation}
The force can thus be expressed as:
\begin{equation}
F_x = C \cos(2\theta+\Psi)
\end{equation}
where:
\begin{equation}
C = \frac{9G}{8d^4}\rho_{rot}b(r_{max}^4-r_{min}^4)\sin(\alpha) \rho_{mir} \pi r_{mir}^2 x_{mir} \cos(\phi) (A^2+B^2)^{1/2}
\end{equation}
\begin{equation}
\tan(\Psi) = -\frac{B}{A}
\end{equation}
The amplitude of the supended mirror motion at twice the rotor frequency follows a free test mass response above the resonant frequency of the suspension ($f \gg 0.6~$Hz):
\begin{align}
a(f_{2rot}) & = \frac{C}{M(2\pi f_{2rot})^2} \nonumber \\
&~ = \frac{9G}{32 \pi^2 f_{2rot}^2 d^4}\rho_{rot}b(r_{max}^4-r_{min}^4)\sin(\alpha) \cos(\phi) (A^2+B^2)^{1/2}
\label{eq:app20}
\end{align}
One can notice that when the center of the rotor is in the plane of the interferometer ($z=0$), the phase shift of the force is $\Psi = 0$ and the amplitude of the force is $C=A$. The amplitude of the mirror motion at twice the rotor frequency is eventually:
\begin{align}
a(f_{2rot}) & = \frac{9 G \rho_{rot}~ b~\sin(\alpha)(r_{max}^{4}-r_{min}^{4})}{32\pi^2 f_{2rot}^{2}d^4}\cos(\phi)\Big[1+\frac{25}{54d^2}\frac{(r_{max}^{6}-r_{min}^{6})}{(r_{max}^{4}-r_{min}^{4})}+\Big(\frac{45}{8}\sin^2(\phi)-\frac{5}{2}\Big)\Big(\frac{r_{mir}}{d}\Big)^2 \nonumber \\
&~~~~~~~~~~~~~~~~~~~~~~~~~~~~~~~~~~~~~~~~~~~~~~~~~~~~~+\Big(\frac{15}{8}\cos^2(\phi)-\frac{25}{24}\Big)\Big(\frac{x_{mir}}{d}\Big)^2-\frac{25}{72}\Big(\frac{b}{d}\Big)^2\Big]
\label{eq:app21}
\end{align}

\newpage

\end{document}